\documentstyle[12pt]{article}

\newcommand{\be}{\begin{equation}}
\newcommand{\ee}{\end{equation}}
\newcommand{\Dlt}{\Delta}
\newcommand{\dlt}{\delta}
\newcommand{\prt}{\partial}
\newcommand{\brr}{{\bf r}}

\newcommand{\bt}{\beta}
\newcommand{\vp}{\varphi}
\newcommand{\ep}{\varepsilon}
\newcommand{\al}{\alpha}
\newcommand{\ra}{\rightarrow}
\newcommand{\sgm}{\sigma}

\newcommand{\gm}{\gamma}
\newcommand{\om}{\omega}

\newcommand{\dgr}{\dagger}
\newcommand{\lbd}{\lambda}

\begin{document}
\begin{center}
{\Large{\bf Entanglement production with Bose atoms in 
optical lattices} \\ [5mm]
V.I. Yukalov$^1$ and E.P. Yukalova$^2$} \\ [3mm]

{\it $^1$Bogolubov Laboratory of Theoretical Physics, \\
Joint Institute for Nuclear Research, Dubna 141980, Russia \\ [2mm]

$^2$Department of Computational Physics, Laboratory of Information
Technologies,\\
Joint Institute for Nuclear Research, Dubna 141980, Russia}

\end{center}

\vskip 3cm

\begin{abstract}
A method of entanglement production is suggested, based on the
resonant generation of topological modes in systems with Bose-Einstein
condensates trapped in optical or magnetic lattices. The method makes
it possible to regulate the strength of entanglement production as well
as to govern its time variation. This method can serve as a practical
tool for quantum information processing and quantum computing.
\end{abstract}

\newpage

\section{Introduction}

Systems with Bose-Einstein condensate (BEC) possess a number of
unusual properties (see book [1] and review articles [2--7]), which
can be employed in a variety of applications. One of such important
applications is the possibility of creating massive entanglement in
Bose-condensed systems trapped in lattice potentials [8,9]. The most
known examples of the lattices are the optical lattices formed by a
combination of laser beams [10--12], though magnetic fields could
also be used for creating magnetic lattices. In experiments, one can
form optical lattices of varying spacing, depth, and different filling
factors, ranging between one and $10^4$ atoms in each lattice site
[13,14]. When a lattice is sufficiently deep and the filling factor
is large, then a lattice site represents a microtrap, in which at low
temperature BEC can be formed.

A BEC in a trap possesses a set of discrete or quasidiscrete atomic
energy levels. If different trapping sites are completely separated
from each other, the energy spectrum is purely discrete. When atoms
can tunnel between the sites, then a line of the discrete spectrum
widens into a band. In what follows, we consider the situation, when
the linewidths are not too large, so that the bands are always well
separated from each other, that is, when the linewidths are much
smaller than the spectrum gaps. The latter situation corresponds to
a quasidiscrete spectrum. Note that we are talking here about the
atomic BEC energy levels in a trap, which should not be confused with
the spectrum of elementary excitations.

In an equilibrium system, BEC sets in the lowest energy level. But
if the system is subject to an alternating modulating field, with
a frequency in resonance with one of the transition frequencies, then
a nonground-state BEC can be realized, as was first proposed in Ref.
[15]. The condensate functions, describing the standard ground-state
BEC and the nonground-state condensates have different spatial
shapes, because of which the condensate wave functions, pertaining
to different energy levels, can be termed topological modes. The
properties of these modes have been theoretically investigated in
a series of papers [15--35] and a dipole topological mode was
generated in experiment [36]. A simple example of such a mode is
a vortex that can be excited in a rotating BEC.

Dynamics of BEC are usually considered in the frame of the 
Gross-Pitaevskii equation (see Refs. [1,2]), which presupposes 
the case of zero temperature and very weak atomic interactions. The 
Gross-Pitaevskii equation was also the basis for describing the resonant 
excitation of topological modes [15]. In the present paper, we show that
topological modes can be created as well at {\it finite temperature},
which is due to the {\it resonant mechanism} of their generation.

Another our goal is to demonstrate how the regulated generation of
topological modes in {\it optical lattices} can be used for the {\it
controlled entanglement production}. The possibility of effectively
varying the state of a complex quantum system and of controlling its
entanglement are the two key points for realizing quantum information
processing and quantum computing [37--42].

\section{Topological modes}

In order to give a general correct definition of topological
coherent modes, we need, first, to write down the exact equation for
the condensate wave function. The latter, for a Bose-condensed system,
is introduced by means of the Bogolubov shift [43--45] for the field
operator
\be
\label{1}
\psi(\brr,t) \; \ra \; \hat\psi(\brr,t) \equiv \eta(\brr,t) +
\psi_1(\brr,t) \; ,
\ee
in which $\eta(\brr,t)$ is the condensate wave function and
$\psi_1(\brr,t)$ is the field operator of uncondensed particles.
The Bogolubov shift (1) explicitly breaks the gauge symmetry of
the Bose system. It is worth emphasizing that the gauge symmetry
breaking is the necessary and sufficient condition for the
occurrence of BEC [46]. After introducing the Bogolubov shift (1),
it is necessary to resort to a representative statistical ensemble
for the Bose system with broken gauge symmetry [47,48], defining
the appropriate grand Hamiltonian $H[\hat\psi]=H[\eta,\psi_1]$.
The theory of Bose-condensed systems is self-consistent only with
the correctly defined grand Hamiltonian [49--51]. The general
equation for the condensate wave function has the form
\be
\label{2}
i \; \frac{\prt\eta(\brr,t)}{\prt t}\;  = \; <
\frac{\dlt H[\eta,\psi_1]}{\dlt\eta^*(\brr,t)} > \; .
\ee

Assuming the standard energy Hamiltonian with the local interaction
potential
\be
\label{3}
\Phi(\brr) = \Phi_0 \dlt(\brr) \; , \qquad
\Phi_0 \equiv 4\pi\; \frac{a_s}{m} \; ,
\ee
where $a_s$ is the scattering length, we obtain from Eq. (2) the 
{\it exact} equation for the condensate wave function
$$
i\; \frac{\prt}{\prt t}\; \eta(\brr,t) = \left ( - \;
\frac{\nabla^2}{2m} + U - \mu_0 \right ) \eta(\brr,t) +
$$
\be
\label{4}
+ \Phi_0 \left \{ \; \left [ \; \rho_0(\brr,t) +
2\rho_1(\brr,t)\; \right ] \eta(\brr,t) +
\sgm_1(\brr,t) \eta^*(\brr,t) + \xi(\brr,t)\; \right \} \; .
\ee
Here $U=U(\brr,t)$ is an external potential and the notation is used
for the condensate density
\be
\label{5}
\rho_0(\brr,t) \equiv |\eta(\brr,t)|^2 \; ,
\ee
the density of uncondensed atoms
\be
\label{6}
\rho_1(\brr,t) \; \equiv \; < \psi_1^\dgr(\brr,t) \psi_1(\brr,t) > \; ,
\ee
the anomalous average
\be
\label{7}
\sgm_1(\brr,t) \; \equiv \; < \psi_1(\brr,t)\psi_1(\brr,t) > \; ,
\ee
and the triple anomalous average
\be
\label{8}
\xi(\brr,t) \; \equiv \; < \psi_1^\dgr(\brr,t) \psi_1(\brr,t)
\psi_1(\brr,t) > \; .
\ee

Looking for the stationary solutions of Eq. (4) in the common form
$$
\eta_n(\brr,t) = \eta_n(\brr) e^{-i\om_n t} \; ,
$$
we come to the stationary equation for the condensate wave function
$$
\left [ -\; \frac{\nabla^2}{2m} + U(\brr) \right ] \eta_n(\brr) +
$$
\be
\label{9}
+ \Phi_0 \left \{ \; \left [\; |\eta_n(\brr)|^2 + 2\rho_1(\brr)\;
\right ] \eta_n(\brr) + \sgm_1(\brr)\eta_n(\brr) +
\xi(\brr)\; \right \} = E_n \eta_n(\brr) \; ,
\ee
in which
\be
\label{10}
E_n \; \equiv \; \mu_0 + \om_n \; .
\ee
In equilibrium, BEC corresponds to the lowest energy level
$$
E_0 \; \equiv \; \min_n E_n = \mu_0 \; , \qquad
\om_0 \; \equiv \; \min_n \om_n = 0  \; .
$$
But, generally, the nonlinear eigenvalue problem (9) possesses
a set of energy levels $E_n$ and of the related eigenfunctions
$\eta_n(\brr)$, labelled by a multi-index $n$. The solutions
$\eta_n(\brr)$ to the eigenvalue problem (9) are the {\it topological
coherent modes}. It is only in the limiting case of zero temperature
and asymptotically weak interactions, when we can neglect $\rho_1(\brr)$, 
$\sgm_1(\brr)$, and $\xi(\brr)$ in Eq. (9), we come to the stationary 
Gross-Pitaevskii equation
$$
\left [ -\; \frac{\nabla^2}{2m} + U(\brr) + \Phi_0 |\eta_n(\brr)|^2
\right ]\eta_n(\brr) = E_n\; \eta_n(\brr) \; ,
$$
where, actually, solely the ground-state level $E_0=\mu_0$ is to be
considered [1].

\section{Resonant generation}

Equation (9) defines the topological modes as stationary solutions.
In order to realize transitions between modes, it is necessary to
include a time-dependent external potential and to consider the
temporal equation (4).

Suppose that at the initial time $t=0$ the system is in equilibrium
and the condensate wave function corresponds to the standard
ground-state condensate,
\be
\label{11}
\eta(\brr,0) = \eta_0(\brr) \equiv \eta(\brr) \; .
\ee
Let us wish to generate a topological mode labelled by the index
$n=n_1$, with the related energy $E_1\equiv E_{n_1}$. Hence, the
transition frequency is given by
\be
\label{12}
\om_{10} \equiv E_1 - E_0 = E_{n_1} - \mu_0 \; .
\ee
To generate this mode, it is necessary to apply an alternating field
\be
\label{13}
V(\brr,t) = V_1(\brr)\cos(\om t) + V_2(\brr) \sin(\om t)
\ee
with a frequency $\om$ tuned close to the transition frequency (12),
so that the resonance condition
\be
\label{14}
\left | \frac{\Dlt\;\om}{\om} \right | \ll 1 \qquad
(\Dlt\; \om \equiv \om - \om_{10} )
\ee
be valid. Then the total external potential in Eq. (4) is the sum
\be
\label{15}
U(\brr,t) = U(\brr) + V(\brr,t)
\ee
of a trapping potential and of the alternating field (13).

The condensate wave function is normalized to the total number of
condensed atoms
\be
\label{16}
N_0 = \int | \eta(\brr,t) |^2 d\brr =
\int |\eta(\brr)|^2 d\brr \; .
\ee
Let us introduce a function $\vp_n(\brr)$, defined by the equality
\be
\label{17}
\eta_n(\brr) \equiv \; \sqrt{N_0}\; \vp_n(\brr) \; ,
\ee
which is normalized to one
\be
\label{18}
\int |\vp_n(\brr) |^2 \; d\brr = 1 \; .
\ee

We shall look for the solution of Eq. (4) in the form
\be
\label{19}
\eta(\brr,t) = \sum_n \; C_n(t)\; \eta_n(\brr)\; e^{-i\om_n t} \; ,
\ee
where $\om_n\equiv E_n-E_0$ is in agreement with Eq. (10). The
coefficient function $C_n(t)$ is treated as a slow function of time,
such that
\be
\label{20}
\frac{1}{\om_n} \left | \frac{d C_n}{dt} \right | \ll 1 \; .
\ee
This allows us to consider $C_n$ as a quasi-integral of motion and
to employ the averaging method [52] and the scale separation approach
[53,54]. For instance, substituting form (19) into the normalization
condition (16), and averaging the latter over time, with $C_n$ kept
as quasi-integrals, we get
\be
\label{21}
\sum_n \; |C_n(t)|^2 =  1 \; .
\ee
The quantity
\be
\label{22}
p_n(t) \equiv | C_n(t)|^2
\ee
plays the role of the mode probability, or fractional mode population,
which is normalized to one, according to Eq. (21).

Substituting expansion (19) into Eq. (4), we use the averaging
techniques [52--54]. The resonant field (13) can be written as
\be
\label{23}
V(\brr,t) = \frac{1}{2} \left [\; B(\brr) e^{i\om t} +
B^*(\brr) e^{-i\om t}\; \right ] \; ,
\ee
where
\be
\label{24}
B(\brr) \equiv V_1(\brr) - i V_2(\brr) \; .
\ee
We need the notation for the interaction amplitude
\be
\label{25}
\al_{mn} \equiv N_0 \Phi_0 \int |\vp_m(\brr)|^2 \left [ \;
2 |\vp_n(\brr)|^2 - |\vp_m(\brr)|^2 \; \right ]\; d\brr \; ,
\ee
pumping-field amplitude
\be
\label{26}
\bt_{mn} \equiv \int \vp_m^*(\brr) B(\brr) \vp_n(\brr) \;
d\brr \; ,
\ee
where $B(\brr)$ is given by Eq. (24), and for the quantity
\be
\label{27}
\gm_{nn} \equiv \al_{nn} - \Phi_0 \int \vp^*_n(\brr) \left \{
2 \left [\; \rho_1(\brr) - \rho_1(\brr,t)\; \right ] \vp_n(\brr) +
\sgm_1(\brr) \vp_n^*(\brr) + \frac{\xi(\brr)}{\sqrt{N_0}}
\right \} d\brr \; .
\ee
Also, we introduce the effective detuning
\be
\label{28}
\Dlt_{mn} \equiv \Dlt\; \om + \al_{mm} - \al_{nn} \; ,
\ee
in which
$$
\al_{nn} = N_0 \Phi_0 \int |\vp_n(\brr)|^4 \; d\brr \; .
$$

From expression (27) it follows that there exists an effective time
$t_{eff}$, during which $\gm_{nn}$ can be treated as a real quantity,
such that
\be
\label{29}
|{\rm Im}\; \gm_{nn}|\; t_{eff} \ll 1 \; ,
\ee
being weakly dependent on time, in the sense that
\be
\label{30}
\left | \frac{t_{eff}}{\gm_{nn}} \;
\frac{d\gm_{nn}}{dt} \right | \ll 1 \; .
\ee
This effective time is of the order
\be
\label{31}
t_{eff} \sim \frac{1}{\rho_1\Phi_0} \; ,
\ee
where $\rho_1$ is the density of uncondensed atoms. When practically
all atoms are condensed, so that $\rho_1\ra 0$, then $t_{eff}\ra\infty$.
Defining
\be
\label{32}
c_n(t) \equiv C_n(t) \exp(i\gm_{nn} t) \; ,
\ee
we see that, for the times shorter than $t_{eff}$, the fractional mode
population (22) can be written as
\be
\label{33}
p_n(t) \cong |c_n(t)|^2 \qquad ( 0\leq t < t_{eff}) \; .
\ee

The initial condition (11) for the mode amplitude (32) takes the form
\be
\label{34}
c_n(0) = \dlt_{n0} \; .
\ee
With this initial condition, we obtain the equations
$$
i\; \frac{dc_0}{dt} = \al_{01} |c_1|^2 c_0 \; + \; \frac{1}{2}\;
\bt_{01} c_1 e^{i\Dlt_{01} t} \; ,
$$
\be
\label{35}
i\; \frac{dc_1}{dt} = \al_{10} |c_0|^2 c_1 \; + \; \frac{1}{2}\;
\bt^*_{01} c_0 e^{-i\Dlt_{01} t} \; ,
\ee
where $\Dlt_{01}=\Dlt\; \om+\al_{00} -\al_{11}$.

Equations (35) can be simplified by separating the absolute values $|c_n|$
and the phases $\pi_n$ of the complex quantities
\be
\label{36}
c_n = |c_n| e^{i\pi_n t} \; .
\ee
Let us also introduce the following parameters
\be
\label{37}
\al \equiv \frac{1}{2} (\al_{01} + \al_{10} ) \; , \qquad
\bt \equiv |\bt_{01}| = \bt_{01} e^{-i\gm} \; , \qquad
\dlt \equiv \Dlt_{01} + \frac{1}{2} ( \al_{01} - \al_{10} ) \; .
\ee
By defining the population imbalance
\be
\label{38}
s \equiv |c_1|^2 - |c_0|^2
\ee
and the phase difference
\be
\label{39}
x \equiv \pi_1 - \pi_0 + \gm + \Dlt_{01} \; ,
\ee
we can transform Eqs. (35) to the two-dimensional dynamical system
\be
\label{40}
\frac{ds}{dt} = \; - \bt \; \sqrt{1-s^2}\; \sin x \; , \qquad
\frac{dx}{dt} = \al s + \frac{\bt s}{\sqrt{1-s^2}}\; \cos x + \dlt \; .
\ee
Solving these equations defines the fractional mode populations (33) as
$$
p_0(t) = \frac{1-s(t)}{2} \; , \qquad p_1(t) = \frac{1+s(t)}{2} \; .
$$

Equations (35) and (40) were derived earlier [15--17] for a purely
coherent system at zero temperature, when all atoms were in BEC, so that
$N_0=N$. Here we have showed that the same equations can be obtained for
a system at finite temperature, when $N_0<N$. The main difference is that,
when the density of uncondensed atoms is not zero, then Eqs. (35) and (40)
are valid not for all times, but in the time interval $0\leq t<t_{eff}$,
limited by the effective critical time (31).

In the same way, we could derive the equations for the dynamics of
several topological modes, generated by the quasiperiodic modulating field
$$
V(\brr,t) = \frac{1}{2} \; \sum _n \left [
B_n(\brr) e^{i\ep_n t} + B_n^*(\brr) e^{-i\ep_n t} \right ] \; ,
$$
with several frequencies $\ep_n$ tuned to the resonance with different
transition frequencies $\om_{mn}\equiv E_m-E_n$. For example, the
equations for three topological modes would have the form as in Refs.
[31,32], or similar to the equations for three coupled BEC [55]. The
systems with multiple generated topological modes display a variety of
interesting effects, such as interference patterns and interference
currents [20,24], mode locking [15,24,26], dynamical transitions and
critical phenomena [17,20,21,24], chaotic motion [31,32], harmonic
generation and parametric conversion [31,32] that are analogous to these
effects in optics and for elementary excitations in Bose-condensed
systems [56--58], atomic squeezing [24,27,28], which can also be called
spin squeezing, Ramsey fringes [59], and massive entanglement production, 
which, being similar to the entanglement of two atoms, differs from the 
latter by occurring for multiatomic condensates. In the following sections, 
we explain how it is possible to create and regulate entanglement in a 
Bose-condensed system with topological modes in optical lattices.

\section{Coherent states}

First of all, we need to define a basis of states that we shall consider
in what follows. It is convenient to use the basis of coherent states.

Let us consider an optical lattice with $N_L$ sites, each site
representing a deep well with a large filling factor $\nu_j\gg 1$ and with
the number of condensed atoms in a well $N_j$, so that
\be
\label{41}
N_0 = \sum_j N_j \; , \qquad N = \sum_j \nu_j \; ,
\ee
where $j=1,2,\ldots,N_L$. Suppose that each well is subject to the action
of a modulating field $V_j(\brr,t)$ generating topological modes inside
that well. Let a multi-index $n_j$ label the topological modes in the
$j$-th site well, and let $\eta_n(\brr)$ be the related coherent modes,
normalized to the number of condensed atoms in the well,
\be
\label{42}
N_j = \int |\eta_{n_j}(\brr)|^2 d\brr \; .
\ee
For generality, we consider the case when $N_j$ and $\nu_j$ can be
different for differing lattice sites. This can happen, e.g., if the
lattice is perturbed by a disordering potential.

Topological modes are the solutions to nonlinear equations of type (9),
because of which they are not necessarily orthogonal to each other, so
that the scalar product
\be
\label{43}
N_{ij} \equiv \int \eta^*_{n_i}(\brr)\; \eta_{n_j}(\brr) \; d\brr
\ee
is, generally, not zero for $i\neq j$. The diagonal elements of
$N_{jj}=N_j$ are the numbers of condensed atoms (42).

For an $n_j$-mode, we may construct the coherent states in the Fock space
as
\be
\label{44}
|n_j>\; = \left [ \frac{\exp(-N_j/2)}{\sqrt{k!}} \; \prod_{l=1}^k
\eta_{n_j}(\brr_l) \right ] \; ,
\ee
which is a column with respect to $k=0,1,2,\ldots$. The coherent states
(44) are not orthogonal to each other, yielding the scalar product
\be
\label{45}
< n_i|n_j> \; = \; \exp\left ( -\; \frac{N_i+N_j}{2} +
N_{ij} \right ) \; ,
\ee
but each of them is normalized to one, so that $<n_j|n_j>=1$.

Let us define the correlation factor
\be
\label{46}
\lbd_{ij} \equiv \frac{N_{ij}}{\sqrt{N_iN_j}} \; ,
\ee
for which $\lbd_{jj}=1$. From the Cauchy-Schwartz inequality for Eq. (43),
we have
\be
\label{47}
|N_{ij}|^2 < N_i N_j \qquad (i\neq j) \; .
\ee
Hence, for factor (46), we get
$$
| \lbd_{ij} | < 1 \qquad (i\neq j) \; .
$$
The expression in the exponential of the right-hand side of Eq. (45)
can be rewritten by using the equality
$$
\frac{1}{2} (N_i + N_j) - N_{ij} = \frac{1}{2} ( N_i - N_j )^2 +
( 1 - \lbd_{ij} ) N_i N_j \; .
$$
The latter diverges if either $N_i$ or $N_j$, or both, tend to infinity
and $i\neq j$. Thus, we come to the conclusion that the coherent states
(44) are asymptotically orthogonal,
\be
\label{48}
< n_i| n_j > \; \simeq \dlt_{ij} \qquad (N_i + N_j \gg 1) \; .
\ee
They also are asymptotically complete (or overcomplete) in the weak sense,
\be
\label{49}
\sum_{n_j} | n_j><n_j | \simeq 1 \qquad (N_i + N_j \gg 1) \; .
\ee
Therefore, the set $\{ |n_j>\}$ forms a basis, which is asymptotically
orthogonal and complete. The closed linear envelope of this basis forms a
Hilbert space ${\cal H}_j$. The tensor products
\be
\label{50}
| {\bf n} > \; \equiv \; \otimes_j |n_j > \qquad
( {\bf n} \equiv \{ n_j \} )
\ee
compose an asymptotically orthogonal and complete basis $\{|{\bf n}>\}$,
whose closed linear envelope is the Hilbert space ${\cal H}\equiv\otimes_j
{\cal H}_j$. The states of BEC, which is a coherent subsystem of the
physical system, can be interpreted as the vectors of the space ${\cal 
H}$.

\section{Lattice register}

By varying the resonant modulating field acting on the lattice, it is
feasible to govern the creation and behavior of the topological coherent
modes and to regulate entanglement produced in the system. To quantify
the level of the produced entanglement, we shall use the measure of
entanglement production introduced in Ref. [60].

The density operator, characterizing the coherent modes, can be
represented as an expansion over the basis $\{|{\bf n}>\}$,
\be
\label{51}
\hat\rho = \sum_{{\bf n}}\; p_{{\bf n}} | {\bf n} > < {\bf n} | \; ,
\ee
with the normalization
$$
{\rm Tr}_{{\cal H}}\; \hat\rho = \sum_{{\bf n}} p_{{\bf n}} =  1 \; .
$$
Let us define a single-partite operator
\be
\label{52}
\hat\rho_j \equiv
{\rm Tr}_{{\cal H}\setminus{\cal H}_j }\; \hat\rho \; .
\ee
From this definition and Eq. (51), we have
\be
\label{53}
\hat\rho_j = \sum_{{\bf n}}\; p_{{\bf n}}\; | n_j> < n_j | \; .
\ee
Then we define the factor operator
\be
\label{54}
\hat\rho^\otimes \equiv \otimes_j \hat\rho_j \; ,
\ee
for which
$$
{\rm Tr}_{{\cal H}}\; \hat\rho^\otimes =
\prod_j {\rm Tr}_{{\cal H}_j} \; \hat\rho_j = 1 \; .
$$

The measure of entanglement, produced by the density operator (51) is
defined [60] as
\be
\label{55}
\ep(\hat\rho) \equiv \log \;
\frac{||\hat\rho||_D}{||\hat\rho^\otimes||_D} \; ,
\ee
where the logarithm is to the base 2 and $||\cdot||_D$ implies the norm
over the disentangled set
\be
\label{56}
D \equiv \{ f = \otimes_j \vp_j \; | \vp_j\in {\cal H}_j \} \; .
\ee
Here the norms are defined as follows. The set $D$ is assumed to be 
unitary, that is, for any two vectors $f\in D$ and $f'\in D$, one can 
introduce the scalar product $(f,f')$, which is the standard requirement 
for any physical system. Having the scalar product makes it 
straightforward to define the vector norm
$$
||f||_D \; \equiv \; \sqrt{(f,f)} \qquad (f\in D) \; ,
$$
generated by this scalar product. Then, for any linear operator $\hat A$ 
on $D$, the operator norm is given as
$$
||\hat A||_D \; \equiv \; \sup_{f,f'} \; 
\frac{|(f,\hat Af')|}{||f||_D||f'||_D} \qquad
(f\neq 0, \; f'\neq 0) \; .
$$
This can also be represented as
$$
||\hat A||_D \; \equiv \sup_{f,f'} |(f,\hat Af')| \qquad
(||f||_D = ||f'||_D = 1) \; .
$$
Vectors $f\in D$ and $f'\in D$ have the product form similar to 
Eq. (50). 

For the norms of operators (51), (53), and (54), we have
$$
||\hat\rho||_D =\sup_{ {\bf n} } p_{{\bf n} } \; , \qquad
||\hat\rho_j||_{{\cal H}_j} = \sup_{ n_j} \sum_{{\bf n}(\neq n_j)}
 p_{{\bf n} } \; , \qquad
||\hat\rho^\otimes||_D = \prod_j  ||\hat\rho_j||_{{\cal H}_j} \; .
$$
So that for measure (55), we obtain
\be
\label{57}
\ep(\hat\rho) =  \log \; \frac{\sup_{{\bf n}} p_{{\bf n}} }
{\prod_j \sup_{ n_j} \sum_{{\bf n}(\neq n_j)} p_{{\bf n} } } \; .
\ee

Entanglement is generated in the system if and only if
\be
\label{58}
\sup_{{\bf n}} p_{{\bf n}} \neq
\prod_j \sup_{ n_j} \sum_{{\bf n}(\neq n_j)} p_{{\bf n}} \; .
\ee
For example, if all lattice sites would be completely independent, 
such that $p_{{\bf n}}$ would be a product of some $p_{n_j}$, 
$p_{{\bf n}}\ra\prod_j p_{n_j}$. Then, since
$$
\sup_{{\bf n}}  \prod_j p_{n_j} \; \ra \;
\prod_j \sup_{n_j} p_{n_j} \; ,
$$
measure (57) would be zero, $\ep(\hat\rho)\ra 0$, that is, no entanglement
would be produced.

The opposite case would be if all lattice sites were correlated, so that
\be
\label{59}
p_{{\bf n}} = p_n \prod_j \dlt_{n n_j} \; .
\ee
This can be realized if all the lattice is shaken synchronically, with the
same topological mode being generated in all lattice sites. In that case,
$$
\sup_{{\bf n}} p_{{\bf n}} = p_n \; , \qquad
\sum_{{\bf n}(\neq n_j)} p_{{\bf n}} = p_n \dlt_{n n_j} \; ,
$$
because of which
$$
||\hat\rho||_D = p_n \; , \qquad
||\hat\rho^\otimes ||_D = p_n^{N_L} \; .
$$
As a result, the measure of entanglement production (57) becomes
\be
\label{60}
\ep(\hat\rho) = ( 1 - N_L ) \log\; \sup_n\; p_n \; .
\ee
If $M$ topological modes are simultaneously generated in each lattice 
site, then the maximal entanglement production is achieved for $p_n=1/M$.
In such a case, keeping in mind that $N_L\gg 1$, one gets
$$
\ep(\hat\rho) = N_L \; \log\; M \; .
$$
For the two-mode case, this reduces to $\ep(\hat\rho)=N_L\log 2$.
By varying the resonant modulating fields, it is possible to regulate
entanglement in a wide diapason between zero and $N_L\log M$. The
fractional mode populations are given by $p_n=p_n(t)$, which is defined
in Eq. (33).

The resonant process of mode generation is a fast process, occurring on
the time scale $1/\al$. The latter is much shorter than the thermal
effective time (31), provided that the number of condensed atoms $N_0\gg
N_1$ is essentially larger than the number of uncondensed atoms $N_1$.
Another temporal restriction is imposed by the power broadening, defining
the resonance time $t_{res}$, after which nonresonant levels become
excited, even though the modulating field is resonant. The resonance time
can be estimated [24] as
$$
t_{res} = \frac{\al^2\om}{\bt^2(\al^2+\bt^2)} \; .
$$
For $\bt\leq \al$ and $\al\ll\om$, the resonance time
$t_{res}\sim\om/\bt^2$, is much longer that $1/\al$. It looks, therefore,
feasible to achieve sufficiently long decoherence times allowing for the
functioning of the lattice register that can be used for quantum 
information processing and the creation of a boson lattice quantum
computer.

In conclusion, we have shown that the generation of topological coherent 
modes is feasible in Bose systems not only at zero temperature and under 
asymptotically weak interactions, when the whole system would be almost 
completely condensed, but also at finite temperatures and interactions. 
This becomes possible because of the {\it resonant} character of the 
suggested mode generation. An important feature of the resonant mode 
generation is the feasibility of controlling the process, thus, allowing 
one to govern the level of entanglement production realized in an optical 
lattice. Such a possibility of regulating entanglement production in a 
lattice could be employed for creating boson lattice registeres for 
quantum information processing.

\end{document}